\documentclass[10pt, conference, compsocconf]{IEEEtran}
\usepackage{amsmath,amsthm,amsfonts,amssymb,latexsym}

 \usepackage{graphicx}

\newcommand{\beq}{\begin{equation}}
\newcommand{\eeq}{\end{equation}}

\def\ear{\end{array} }

\def\fr1c#1{\frac{1}{#1}}
\def\fr1c#1{\frac{1}{#1}}

\def\text{\rm}

\newtheorem{remark}{Remark}

\ifCLASSINFOpdf
   \usepackage{graphicx}
\else
\fi
\begin{document}
%
\title{Computational Approach to Gravitational Waves Forms in Stellar Systems as Complex Structures through Keplerian Parameters}


\author{\IEEEauthorblockN{Bochicchio Ivana}
\IEEEauthorblockA{Dipartimento di Matematica e Informatica\\
Universit\'{a} degli studi di Salerno\\
Fisciano (SA), Italy\\
and INFN Sez. di Napoli, \\Napoli, Italy\\
Email: ibochicchio@unisa.it}

\and \IEEEauthorblockN{De Laurentis Mariafelicia}
\IEEEauthorblockA{Dipartimento di Scienze Fisiche\\
Universit\'{a} di Napoli `` Federico II'',\\
and INFN Sez. di Napoli, \\
Napoli, Italy\\
Email:felicia@na.infn.it}

\and \IEEEauthorblockN{Laserra Ettore}
\IEEEauthorblockA{Dipartimento di Matematica e Informatica\\
Universit\'{a} degli studi di Salerno\\
Fisciano (SA), Italy\\
and INFN Sez. di Napoli, \\Napoli, Italy\\
Email: elaserra@unisa.it} }


%


\maketitle

\begin{abstract}
In this paper we investigate the gravitational waves emission by
stellar dynamical structures as complex systems in the quadrupole
approximation considering bounded and unbounded orbits. Precisely,
after deriving analytical expressions for the gravitational wave
luminosity, the total energy output and gravitational radiation
amplitude, we present a computational approach to evaluate the
gravitational wave-forms from elliptical, circular, parabolic and
hyperbolic orbits as a function of Keplerian parameters.

\end{abstract}




%
\IEEEpeerreviewmaketitle

\section{Introduction}
In recent years, detailed information has been achieved for
kinematics and dynamics of stars moving in the gravitational field
of such a central object. The statistical properties of spatial
and kinematical distributions are of particular interest. Using
them, it is possible to give a quite accurate estimate of the mass
and the size of the central object. More precisely, in
\cite{ghez}, it is described a campaign of observations where
velocity measurements are extremely accurate. Then from this bulk
of data, considering a field of resolved stars whose proper
motions are accurately known, one can classify orbital motions and
deduce, in principle, the rate of production of GWs according to
the different types of orbits. This motivates this paper in which,
by a classification of orbits in accordance with the conditions of
motion, we want to calculate the GW luminosity for the different
types of stellar encounters. Following the method outlined in
\cite{peters1,peters2,capozziello_de-laurentis}, we investigate
the GW emission by stellar dynamical structures as complex systems
in the quadrupole approximation considering bounded (circular or
elliptical) and unbounded (parabolic or hyperbolic) orbits.
\newline
The main parameter is the approaching energy of the stars in the
system (see also \cite{shutz} and references therein). However, in
a recent work \cite{bo-laGRG}, it is shown that the parametric
form of the main evolution equation of
Lema$\rm{\hat\rm{\i}}$tre-Tolman-Bondi (LTB) cosmological model
can be obtained considering the limiting rectilinear solutions of
Kepler's problem. More precisely, the analogy between the
relativistic evolution of LTB shells and the classical discussion
of a falling body in a Newtonian center of attraction, allows us
to conclude that the relativistic orbits are clearly related to
suitable astronomic coordinates. Here, these coordinates can be
adopted in order to have a detailed investigation of the
gravitational waves emission by stellar encounters.
\newline
Precisely, briefly recalling the main features of stellar
encounters and orbit classification, we present the computational
aspect to obtain the gravitational wave-forms for elliptical,
parabolic and hyperbolic encounters.

\section{Orbit in Stellar Encounters}

Let us take into account the Newtonian theory of orbits since
stellar systems, also if at high densities and constituted by
compact objects, can be usually assumed in Newtonian regime. We
give here a self-contained summary of the well-known orbital types
in order to achieve below a clear classification of the possible
GW emissions. We refer to the text books \cite{binney,landau,roy}
for a detailed discussion.
\newline
A mass $m_1$ is moving in the gravitational potential $\Phi$
generated by a second mass $m_2$. The vector radius and the polar
angle depend on time as a consequence of the star motion, i.e.
\textbf{r} = \textbf{r}(t) and $\phi\,=\,\phi(t)$. As it is known,
the shape of the orbit depends by the total energy $H$ of the
system. Precisely
\par
when ${H}<0$ the orbit is an ellipse and the equation of the
trajectory is
\begin{equation}\label{Parellisse}
r\,=\,\frac{l}{1+\varepsilon \cos\phi} \ ,
\end{equation}
where $l$ is the so-called semi-latus rectum or the parameter of
the ellipse and $\varepsilon$ is the eccentricity of the ellipse
(for more details see \cite{capozziello_de-laurentis}).
\newline
In a generic elliptic Keplerian motion, called $\alpha$ the major
semi–axis and $E$ the eccentric anomaly, the orbit can be written
also as (see \cite{bo-laGRG,roy})
\begin{equation}\label{Ellisse}
r = \alpha (1 - \varepsilon cosE)
\end{equation}
hence, there is the following relation between the eccentric
anomaly and the angle $\phi$:
\begin{equation}\label{camb_E}
\cos \phi\,=\,\frac{\cos E - \varepsilon}{1-\varepsilon \cos E}.
\end{equation}
\par
While ${H}\geq 0$ is the condition to obtain unbounded orbits. The
trajectory is
\begin{equation}\label{ParIpe}
r\,=\,\frac{l}{1+\varepsilon \cos\phi} \ ,
\end{equation}
where $\varepsilon \geq 1$. The equal sign corresponds to $H = 0$.
Therefore, in order to ensure positivity of r, the polar angle
$\phi$ has to be restricted to the range given by
$$1+\varepsilon \cos\phi > 0 .$$
This means $\cos \phi >1$, i.e. $\phi \in (-\pi, \pi)$ and the
trajectory is not closed any more. For $\phi \rightarrow \pm \pi$,
we have $r \rightarrow \infty$. The curve \eqref{ParIpe}, with
$\varepsilon\,=\,1$, is a parabola. For $\varepsilon >1$, the
allowed interval of polar angles is smaller than $\phi \in (-\pi,
\pi)$, and the trajectory is a hyperbola. Such trajectories
correspond to non-returning objects (for more details see
\cite{capozziello_de-laurentis}).
\newline
Let's consider a generic Hyperbolic Keplerian Orbit, with
eccentricity $\varepsilon$, real semi–-axis $\alpha$ and $F$ as
variable, analogous to the elliptic eccentric anomaly $E$. The
orbit is defined by
\begin{equation}\label{Iperbole}
r = \alpha(\varepsilon \cosh F - 1) ;
\end{equation}
hence, there is the following relation between $F$ and the angle
$\phi$:
\begin{equation} \label{camb_F}
\cos \phi\,=\,\frac{l-\alpha(\varepsilon \cosh F - 1)}{\varepsilon
\alpha (\varepsilon \cosh F - 1)}.
\end{equation}
Let's consider the Keplerian Parabolic Orbit. It is defined by the
relation
\begin{equation}\label{Parabola}
r \,=\, \frac{P^2}{2}
\end{equation}
where P is a parameter. In this case
\begin{equation} \label{camb_P}
\cos \phi \,=\, \frac{2 l - P^2}{P^2} .
\end{equation}
\par
In the next section we present the \texttt{computational approach}
 to obtain gravitational wave--forms in the different cases as
function of the eccentric anomaly, the angle $F$ and the parameter
$P$.
\begin{remark}
Let the eccentricity tend to unity in equations
\eqref{Ellisse}--\eqref{Parabola}; in the limit the corresponding
orbits are called rectilinear ellipse, hyperbola and parabola
respectively.
\newline
The evolution of the r–-shell in LTB models is analogous to a
Keplerian motion on a rectilinear ellipse, hyperbola or parabola
(for more details see \cite{bo-laGRG}).
\end{remark}

\section{Computational Approach to the Gravitational Wave Forms}

At this point, considering the orbit equations, we want to
classify the gravitational radiation for the different stellar
encounters. We send the Reader to the References \cite{misner},
\cite{shapiro}, \cite{maggiore}, \cite{thorne} for a detailed
exposition.
\newline
The Einstein field equations give a description of how the
curvature of space-time is related to the energy-momentum
distribution. In the weak field approximation, moving massive
objects produce gravitational waves which propagate in the vacuum
with the speed of light. One can search for wave solutions of
field equation from a system of masses undergoing arbitrary
motions, and then obtain the power radiated
\cite{capozziello_de-laurentis}. The result, assuming the source
dimensions very small with respect to the wavelengths (quadrupole
approximation \cite{landau}), is that the power
$\frac{dH}{d\Omega}$ radiated in a solid angle $\Omega$ with
tensor of polarization $e_{ij}$ is
$$\frac{dH}{d\Omega}\,=\, \frac{G}{8 \pi c^5}\left(\frac{d^3Q_{ij}}{dt^3}e_{ij}\right)$$
where $Q_{ij}$ is the quadrupole mass tensor
$$Q_{ij} = \sum_{a} m_a(3x^i_a x^j_a - \delta_{ij} r^2_a)$$
$G$ being the Newton constant, $r_a$ the modulus of the vector
radius of the a-th particle and the sum running over all masses
$m_a$ in the system. With this formalism, it is possible to
estimate the amount of energy emitted in the form of GWs from a
system of massive objects interacting among them
\cite{peters1,peters2}. In this case, the components of the
quadrupole mass tensor in the equatorial plane
$(\theta\,=\,\frac{\pi}{2})$ are:
\begin{equation}
\begin{array}{lll}
Q_{xx}=\mu r^2(3\cos{^2\phi}-1)~,\\ \\
Q_{yy}=\mu r^2(3\sin{^2\phi}-1)~,\\ \\
Q_{zz}=-\mu r^2~,\\ \\
Q_{xz}=Q_{zx}=0~,\\ \\
Q_{yz}=Q_{zy}=0~,\\ \\
Q_{xy}=Q_{yx}=3\mu r^2 \cos\phi \sin\phi~,
\end{array}\label{eq:quadrupoli}
\end{equation}
where $\mu=\frac{m_1 m_2}{m_1+m_2}$ is the reduced mass of the
system and the masses $m_1$ and $m_2$ have polar coordinates
$\{r_i \cos \theta \cos \phi, r_i \cos \theta sin \phi, r_i \sin
\theta\}$ with $i = 1, 2$. The origin of the motions is taken at
the center of mass.

Direct signatures of gravitational radiation are its amplitude and
its wave-form. In other words, the identification of a GW signal
is strictly related to the accurate selection of the shape of
wave-forms by interferometers or any possible detection tool. Such
an achievement could give information on the nature of the GW
source, on the propagating medium, and , in principle, on the
gravitational theory producing such a radiation
\cite{cap-delau-franc}. It is well known that the amplitude $h$ of
GWs can be evaluated by
\begin{equation} \label{ampiezza}
h_{jk}(t,R) = \frac{2G}{Rc^4} \ddot{Q}_{jk},
\end{equation}
$R$ being the distance between the source and the observer and
$\{j, k\} = 1, 2$.
\newline
Moreover, in accordance to
\begin{equation} \label{strainAmplitude}
h\simeq(h_{11}^2+h_{22}^2+2h_{12}^2)^{1/2}
\end{equation}
we can compute the expected strain amplitude for different orbits.
\newline
Note that we are going to study the evolution of  compact binary
systems that are formed through the  capture of a moving mass
$m_1$ by a gravitational field, whose source is a Black Hole(BK)
of mass $m_2$ and being $m _1\ll m_2$. So we will have that the
reduced mass of the newborn binary system $\mu\cong m_1$ and the mass ratio $\frac{m_1}{m_2}$ is
$\ll 1$. This constraint is satisfied by several real systems.

Let us now derive the GW amplitude in relation to the orbital
shape of the binary systems.

\subsection{Quadrupole and GW Amplitude from Elliptical Orbits}
In this section the components of the quadrupole mass tensor in
the equatorial plane for the elliptical orbits are expressed.
Precisely, by Eq.s \eqref{Ellisse}, \eqref{camb_E} and
\eqref{eq:quadrupoli}, we obtain:
\begin{equation}\label{eq:quadrupoliEllittico}
\begin{array}{lll}
Q_{xx}=-\alpha ^2 m_1 \left[-3 \varepsilon^2+4\varepsilon \cos E
+\left(\varepsilon^2-3\right) \cos^2 E+1\right],\\\\
Q_{yy}=\alpha^2 m_1 \left[-3 \varepsilon^2+2 \varepsilon \cos E
   +\left(2 \varepsilon ^2-3\right) \cos^2 E+2\right],\\\\
Q_{zz}\,=\,-\alpha ^2 m_1  (\varepsilon  \cos E -1)^2~,\\\\
Q_{xy}\,=\,Q_{yx}\,=& \\
 \qquad \, =\,3 \alpha^2 m_1 (\varepsilon -\cos E) (\varepsilon  \cos
   E-1) \sqrt{-\frac{\left(\varepsilon ^2-1\right) \sin
   ^2 E}{(\varepsilon  \cos E-1)^2}}
\end{array}
\end{equation}
Moreover, considering a binary system and the single components of
Eq. \eqref{ampiezza}, it is straightforward to show that
\begin{equation}
\begin{array}{llllllll}
h_{11}=\frac{\alpha G_N m_2  m_1 \left[(\varepsilon
-1)(\varepsilon +3) \csc^2\left(\frac{E}{2}\right)+2
\left(\varepsilon ^2-3\right)
   \cos E\right]}{\cos E-1}~,
\\ \\
h_{22}=\frac{\alpha G_N m_2 m_1 \left[\left(2 \varepsilon
^2+\varepsilon -3\right) \csc^2\left(\frac{E}{2}\right)+\left(4
\varepsilon^2-6\right) \cos E\right]}{\cos E-1}~,
\\ \\
h_{12}=h_{21}= 3 G_N m_2 m_1 \alpha \sqrt{-\frac{\left(\epsilon^2-1\right) \sin^2E}{(\epsilon  \cos E-1)^2}}\, \cdot \\
\\
\qquad \, \frac{ (\epsilon  \cos E-1)
   (\epsilon -4 \cos E+\cos2E+2) \cot^2\left(\frac{E}{2}\right)
   \csc^2E }
   {\cos E-1}~,
\end{array}
\end{equation}
and then the expected strain amplitude $h$, which strictly depends
on the initial conditions of the stellar encounter, can be
evaluated in accordance to \eqref{strainAmplitude}.
\begin{figure}[!t]
\centering
\includegraphics[width=3.8in]{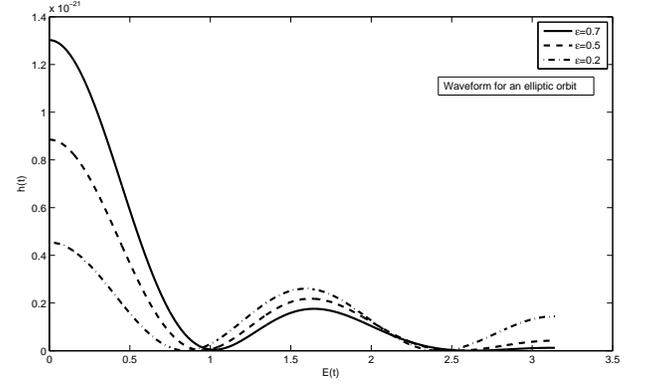}
\caption{The gravitational wave-forms from elliptical orbits shown
as  function of  $E(t)$. We have fixed the masses of the order
$1.4M_{\odot}$. $m_{2}$ is considered at rest while $m_{1}$ is
moving. The distance of the GW source is assumed to be $R=8$ kpc
and the eccentricity is  $\varepsilon= 0.2,0.5, 0.7.$ }
\label{fig_ellisse}
\end{figure}

\subsection{Quadrupole and GW Amplitude from Hyperbolic Orbits}
When hyperbolic orbits are taken into account, Eq.s
\eqref{Iperbole}, \eqref{camb_F} and \eqref{eq:quadrupoli} become:
\begin{equation}\label{eq:quadrupoliIperbolico}
\begin{array}{lll}
Q_{xx}=\alpha^2 m_1 (\epsilon  \cosh F-1)^2 \left[\frac{3
(l+\alpha-\alpha  \epsilon  \cosh F)^2}{\left(\alpha \epsilon ^2
\cosh F-\alpha  \epsilon \right)^2}-1\right] \\\\
Q_{yy}=\alpha^2 m_1 (\epsilon  \cosh F-1)^2 \left[3
   \left(1-\frac{(l+\alpha -\alpha  \epsilon  \cosh F)^2}{\left(\alpha  \epsilon ^2 \cosh F-\alpha  \epsilon
   \right)^2}\right)-1\right]
\\\\
Q_{zz}\,=\,-\alpha ^2 \mu  (\epsilon  \cosh F-1)^2\\\\
Q_{xy}\,=\,Q_{yx}\,= \\
\\
\qquad \, = \frac{3 \alpha  \mu  (\epsilon \cosh F-1) (l+\alpha
-\alpha
   \epsilon  \cosh F) \sqrt{1-\frac{(l+\alpha -\alpha  \epsilon
   \cosh F)^2}{\left(\alpha  \epsilon ^2 \cosh F-\alpha
   \epsilon \right)^2}}}{\epsilon }
\end{array}
\end{equation}

In this case the single components of Eq. \eqref{ampiezza} for a
hyperbolic orbit, are:
\begin{equation}
\begin{array}{llllllll}
h_{11}=\frac{2 G_Nm_2 m_1}{\varepsilon ^2} \left[\frac{3
l^3}{\alpha ^2 (\varepsilon  \cosh F-1)^3}-\frac{9 l^2}{\alpha
(\varepsilon \cosh(F(t))-1)^2}+ \right.
\\
\left.\qquad \quad +\frac{\left(9-2 \varepsilon ^2\right)
l}{\varepsilon \cosh F-1}+\alpha  \left(-2 \varepsilon ^4+8
\varepsilon
   ^2-3\right)\right]~,
\\ \\
h_{22}=\frac{2 G_N m_2 m_1}{\varepsilon ^2} \left[-\frac{3
l^3}{\alpha ^2 (\varepsilon  \cosh F-1)^3}+ \frac{9 l^2}{\alpha
(\varepsilon \cosh F-1)^2} \right. \\
\left. \qquad \quad+\frac{\left(4 \varepsilon ^2-9\right)
l}{\varepsilon \cosh F-1}+ \alpha  \left(4 \varepsilon ^4-10
\varepsilon^2+3\right)\right]~,
\\ \\
h_{12}=h_{21} =-\frac{2G_Nm_2m_1\left(\frac{(l+\alpha -\alpha  \varepsilon  \cosh F)^2}{\alpha ^2 \varepsilon ^2 (\varepsilon  \cosh
   F-1)^2}+1\right)^{3/2}}{\alpha  \varepsilon  (\varepsilon  \cosh F-1)^2}
\cdot \\
\\
\ l^2+\alpha ^2 \left(\varepsilon ^2-1\right)+\alpha ^2
\varepsilon  \left(\varepsilon ^2-1\right) \cosh F (\varepsilon
\cosh (F-2))~,
   \end{array}
\end{equation}
which, as before, strictly depends on the initial conditions of
the stellar encounter.

\begin{figure}[!t]
\centering
\includegraphics[width=3.7in]{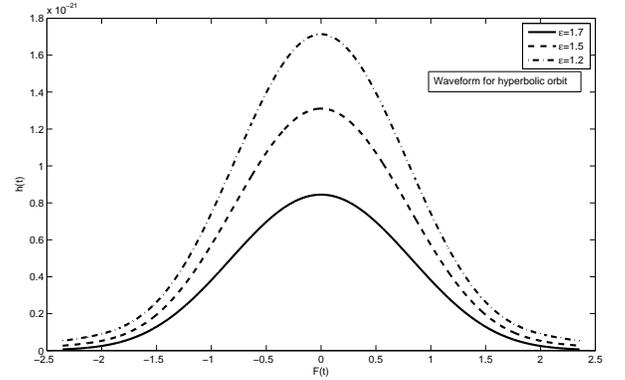}
\caption{The gravitational wave-forms for hyperbolic encounters as
function of the  $F(t)$.  As above, we have fixed the masses of
the order $1.4M_{\odot}$. $m_{2}$ is considered at rest while
$m_{1}$ is moving. The distance of the source is assumed at $R=8$
kpc. The eccentricity is assumed with the values
$\varepsilon=1.2,1.5,1.7$ .} \label{fig_iperbole}
\end{figure}

\subsection{Quadrupole and GW Amplitude from Parabolic Orbits}
Finally, considering parabolic orbits, the explicit expression of
the components of the quadrupole mass tensor in the equatorial
plane through the parameter $P$ are deduced by \eqref{Parabola},
\eqref{camb_P} and \eqref{eq:quadrupoli}:
\begin{equation}
\begin{array}{lll}
Q_{xx}=\frac{1}{4} \left[\frac{3 (2 l-1)^2}{P^4}-1\right] P^4\,m_1~,\\ \\
Q_{yy}=\frac{1}{4} P^4 \left\{\frac{3 \left[P^4-(2 l-1)^2\right]}{P(t)^4}-1\right\} m_1~,\\ \\
Q_{zz}=-\frac{P^4 m_1 }{4}~,\\ \\
Q_{xz}=Q_{zx}=0~,\\ \\
Q_{yz}=Q_{zy}=0~,\\ \\
Q_{xy}=Q_{yx}=\frac{3}{4} (2 l-1) P^2 \sqrt{\frac{P^4-(2
l-1)^2}{P^4}} m_1~,
\end{array}\label{eq:quadrupoliParabolico}
\end{equation}
Moreover, the single components of Eq. \eqref{ampiezza} for a
parabolic orbit, are:
\begin{equation}
\begin{array}{llllllll}
h_{11}=-\frac{4 G_N m_2 m_1 }{P^2}~,
\\ \\
h_{22}=\frac{8 G_N m_2 m_1 }{P^2}~,
\\ \\
h_{12}=h_{21} =\frac{6 G m_2 m_1  \left[(1-2 l) P^4-(2
   l-1)^3\right]}{\left[1-\frac{(1-2 l)^2}{P^4}\right]^{3/2} P^8}~,
\end{array}
\end{equation}
and then the expected strain amplitude $h$, which strictly depends
on the initial conditions of the stellar encounter, can be
evaluated in accordance to \eqref{strainAmplitude}.

\begin{figure}[!t]
\centering
\includegraphics[width=3.7in]{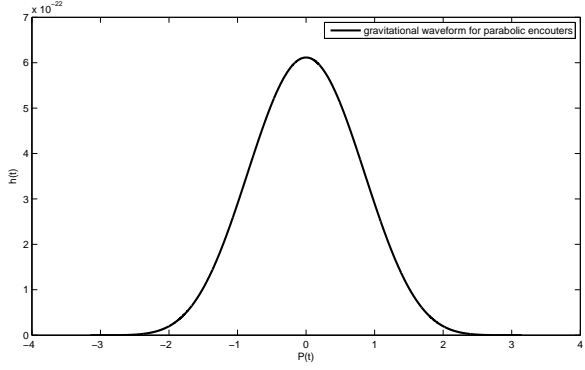}
\caption{The gravitational wave-forms for a parabolic encounter as
a function of  $P(t)$. As above, we have fixed the masses of the
order $1.4M_{\odot}$ and $m_{2}$ is considered at rest, while
$m_{1}$ is moving. The distance of the GW source is assumed at
$R=8$ kpc. The eccentricity is $\varepsilon=1$. }
\label{fig_parabola}
\end{figure}

\section{Conclusion}
We have analyzed the gravitational wave emission coming from
stellar encounters in Newtonian regime and in quadrupole
approximation. In particular, we have taken into account the
expected  strain amplitude of gravitational
radiation produced in tight impacts where two compact objects  with
masses comparable to the Chandrasekhar limit $(\sim
1.4M_{\odot})$. This choice is motivated by the fact that
ground-based experiments like VIRGO or LIGO expect to detect
typical GW emissions from the dynamics of these objects or from
binary systems composed by them (see e.g. \cite{maggiore}). We
would like to underline that the presented approach provides the
necessary support for the application of computational schemes to
obtain qualitative description of the Gravitational Waves.

\section*{Acknowledgment}

The authors would like to thank Prof. Salvatore Capozziello for
his very useful discussions and suggestions which allowed us to
improve the paper.



%

\end{document}